 \documentclass[final,5p,times,twocolumn,authoryear]{elsarticle}

\usepackage{amssymb}
\usepackage{lipsum}
\usepackage{amsmath}
\usepackage{mathtools}
\usepackage{siunitx}
\usepackage{hyperref}
\usepackage{orcidlink}

\journal{Nuclear Physics B}

\begin{document}

\begin{frontmatter}

\title{DARWIN/XLZD: a future xenon observatory for dark matter and other rare interactions}

\author{Laura Baudis\,\orcidlink{0000-0003-4710-1768}\,}
\affiliation{organization={Department of Physics, University of Zurich},
            addressline={Winterthurerstr. 190}, 
            city={Zurich},
            postcode={8057}, 
            country={Switzerland}}

\begin{abstract}
The DARWIN/XLZD experiment is a next-generation dark matter detector with a multi-ten-ton liquid xenon time projection chamber at its core.  Its principal goal will be to explore the experimentally accessible parameter space for Weakly Interacting Massive Particles (WIMPs) in a wide mass-range, until interactions of astrophysical neutrinos will become an irreducible background. The prompt scintillation light and the charge signals induced by particle interactions in the liquid xenon target will be observed by VUV-sensitive, ultra-low background photosensors. Besides its excellent sensitivity to WIMPs with masses above $\sim$5\,GeV, such a detector with its large mass, low-energy threshold and ultra-low background level will also be sensitive to other rare interactions, and in particular also to bosonic dark matter candidates with masses at the keV-scale. We present the detector concept,  discuss the main sources of backgrounds, the technological challenges and some of the ongoing detector design and R\&D efforts, as well as the large-scale demonstrators. We end by discussing the sensitivity to particle dark matter interactions.
 \end{abstract}

\end{frontmatter}

\section{Introduction}
\label{sec:intro}

There is a vast body of evidence for dark matter (DM) across many length scales in the Universe, but its fundamental nature remains a mystery. Cold dark matter  is one of the foundations of the standard model of cosmology, $\Lambda$CDM, accounting for 26.4\% of the critical density, or 84.4\% of the total matter density~\citep{Planck:2018vyg}. While DM candidates extend over a large range of masses and interaction cross sections, two classes of models stand out: Weakly Interacting Massive Particles (WIMPs) and axions, for these are theoretically well-motivated by open questions in particle physics~\citep{ParticleDataGroup:2022pth}.

Direct DM detection experiments search for rare scatters between a DM particle and an atomic nucleus or for interactions with electrons in various target materials. Operated deep underground to reduce background from cosmic rays, these experiments face two major challenges: the small energies, well below tens of keV and perhaps as low as a few meV,  released in a collision and the ultra-low scattering rates.  Because of the unknown DM interaction cross section, expected rates range from about 10 to much less than 1 event per ton of detector material and year, depending on the DM particle mass. This is an extraordinarily small rate, requiring a low energy threshold, an ultra-low background from radioactivity and a target mass  as large as possible, to maximise the probability of a discovery. 

\section{Liquid Xenon Experiments}
\label{sec:xenon_exp}

\begin{figure}[h]
\centering
\includegraphics[width=0.45\textwidth]{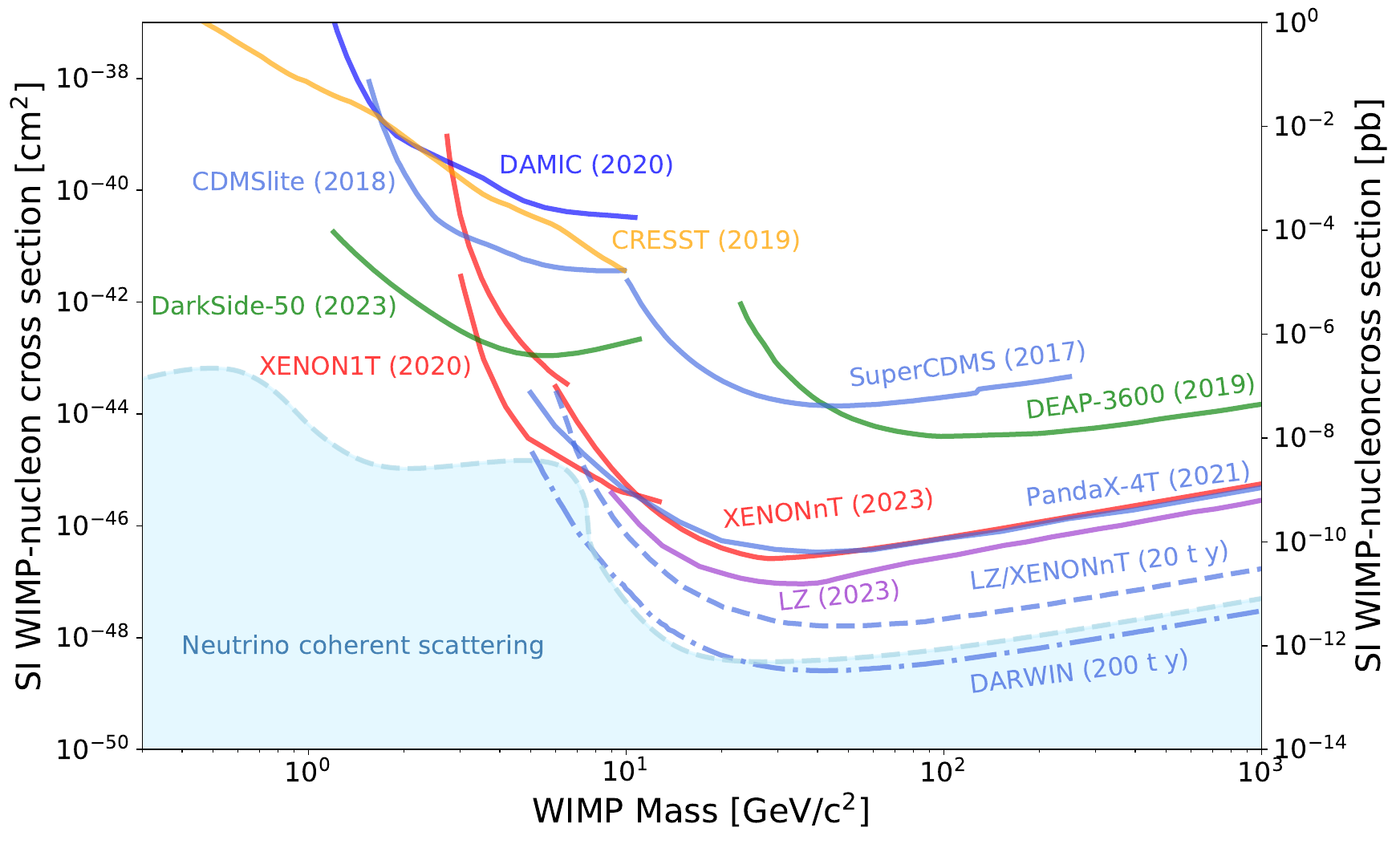}
\caption{Exclusion limits (solid) on the spin-independent WIMP-nucleon cross section from a range of direct DM detection experiments, including dual-phase TPCs. Projections for LZ and XENONnT (dashed) and for DARWIN/XLZD (dashed-dotted) are also shown. The region where a distinction between a dark matter signal and astrophysical neutrinos will be challenging, albeit not impossible~\citep{OHare:2021utq}, is shown in blue.} 
\label{fig:silimits}
\end{figure}

Experiments using liquefied xenon (LXe) as DM target have reached the highest sensitivity to WIMPs with masses above  a few GeV, as shown in Figure~\ref{fig:silimits}. Owing to their low background rates and energy thresholds around 1\,keV, LXe experiments are also sensitive to other new types of particle interactions, as well as to low-energy astrophysical neutrinos. In fact, the latter will induce an irreducible background in the next-generation of detectors at the multi-ten-ton scale. 

Some advantages of xenon experiments are their large and homogeneous detector geometries with efficient self-shielding against external radiation, the fact that their targets are readily purified,  as well as their sensitivity to both, spin-independent and spin-dependent WIMP interactions, given the presence of two isotopes with spin, $^{129}$Xe (26.44\%) and $^{131}$Xe (21.18\%), in natural xenon.   The radioactive isotopes $^{124}$Xe, $^{126}$Xe, $^{134}$Xe and $^{136}$Xe have very long half-lives, and their second-order weak decay modes present interesting physics channels.

\begin{figure}[h]
	\centering 
	\includegraphics[width=0.4\textwidth, angle=0]{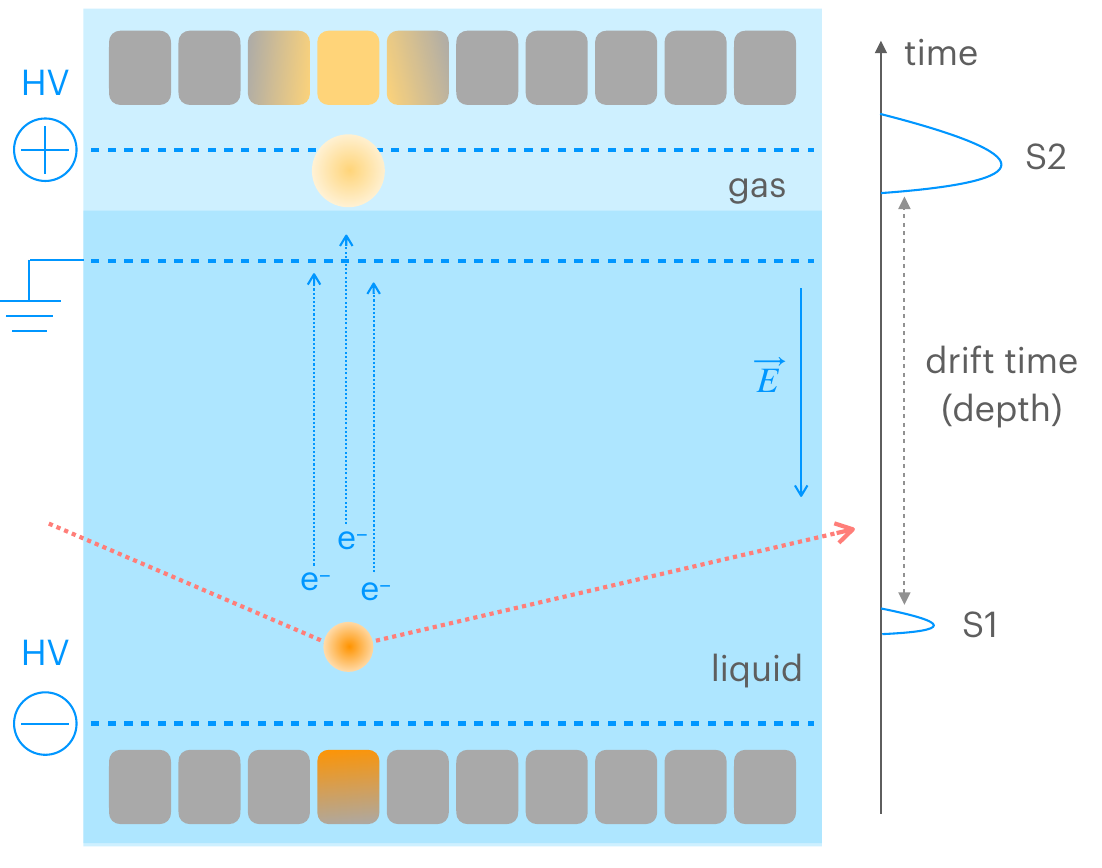}	
	\caption{The operation principle of a dual-phase xenon TPC. A particle interaction in liquid xenon gives rise to a prompt scintillation signal (S1) and a delayed, amplified proportional scintillation signal (S2). The latter is caused by ionisation electrons, which are drifted in a homogeneous electric field (of a few 100\,V/cm) and extracted into the gas phase above the liquid with a higher electric field, typically 10\,kV/cm. The drift field is produced between the cathode at negative potential and a grounded gate grid in the liquid, while the extraction field is obtained by means of the anode placed above the gate in the gas phase. Both S1 and S2 signals are observed with photosensor arrays placed on the bottom and top of the TPC.} 
	\label{fig:tpc}
\end{figure}

Xenon-based experiments employ dual-phase (liquid and gas) TPCs, the working principle of which is shown schematically in Figure~\ref{fig:tpc}. An interaction within the  active volume of a detector will create ionisation electrons and prompt scintillation photons. The prompt scintillation signal (S1) is detected with two arrays of photosensors, one in the liquid phase on the bottom and one in the gas phase at the top. The electrons drift in the pure liquid under the influence of an external electric field, are then accelerated by a stronger field and extracted into the vapour phase above the liquid, where they generate proportional scintillation, or electroluminiscence.  The delayed proportional scintillation signal (S2) is observed by the same photosensor arrays. The array immersed in the liquid collects the majority of the prompt signal, which is totally reflected at the liquid-gas interface. The ratio of the two signals is different for nuclear recoils (NR), such as from fast neutron interactions or  hypothetical {\small WIMPs} and electronic recoils (ER) produced by $\beta$ and $\gamma$-rays, or DM particles scattering off electrons. This provides the basis for background discrimination in dark matter detectors.  Since electron diffusion in the ultra-pure liquid is small (albeit non-negligible), the proportional scintillation photons carry the $x-y$ information of the interaction site. With the $z-$information from the drift time measurement, the {\small TPC} yields  a three-dimensional event localisation, enabling fiducial volume selections and differentiation between single- and multiple-scatters in the active volume~\citep{Baudis:2023pzu}. 

\section{Backgrounds}
\label{sec:backgrounds}

\begin{figure}[!h]
\includegraphics[width=0.45\textwidth]{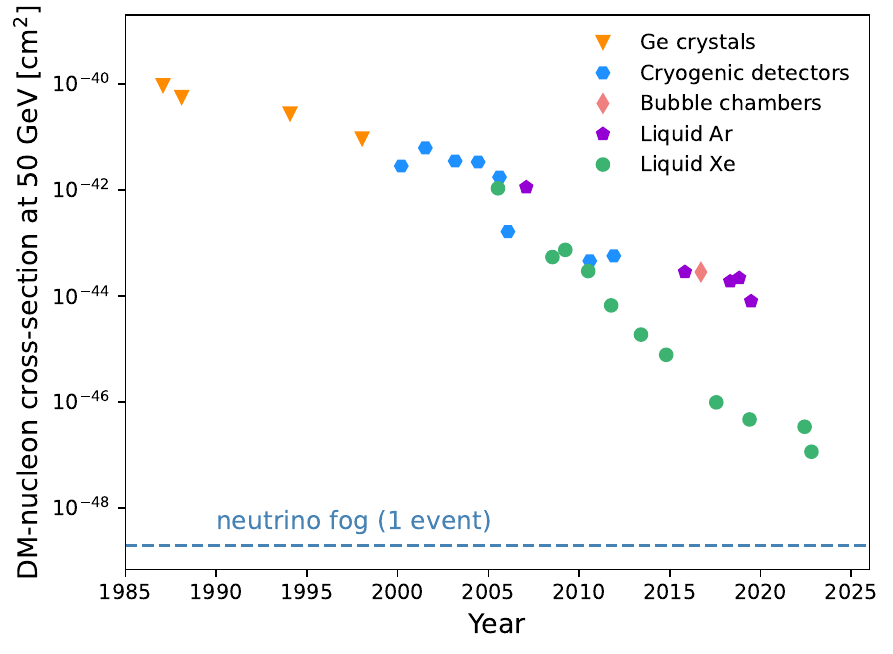}
\caption{Evolution of the sensitivity to spin-independent WIMP-nucleon interactions for a 50 GeV WIMP, for various technologies, including dual-phase Xe-TPCs, with time. The so-called neutrino fog is shown as a horizontal dashed line. Note the logarithmic scale. }
\label{fig:evolution_time}
\end{figure}

The target masses of dual-phase xenon TPCs were gradually scaled up  from a few kg to multi-tons, while the backgrounds were concomitantly reduced for each detector iteration. This lead to the remarkable evolution of the sensitivity to WIMPs, as shown in Figure~\ref{fig:evolution_time}, which covers over three decades.

\begin{figure}[!h]
\includegraphics[width=0.45\textwidth]{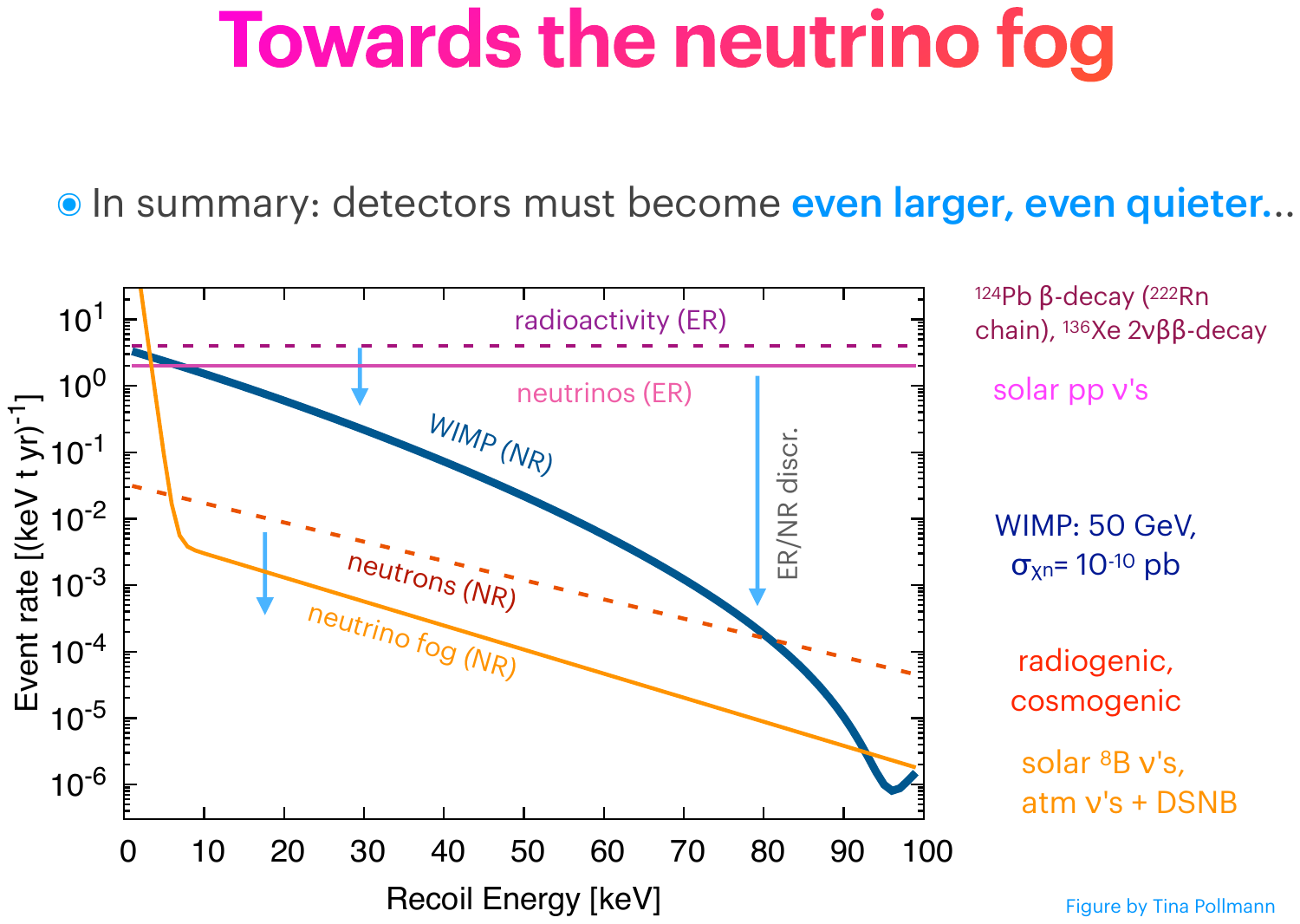}
\caption{Schematic view of the main backgrounds in xenon TPCs, together with a hypothetical, WIMP-induced NR spectrum.  The ER background rates, mainly from radioactivity (radon mixed with the xenon) is to be reduced below the irreducible background from elastic neutrino-electron scatters from solar neutrinos. The NR background rate from neutrons (radiogenic and cosmogenic)  is to be reduced below the irreducible rate from coherent elastic neutrino-nucleus scatters (neutrino fog) from solar and atmospheric neutrinos. Finally ERs and NRs can be distinguished based on their S1 and S2 signals (ER/NR discrimination). Figure by Tina Pollmann.}
\label{fig:bgs}
\end{figure}

\begin{figure}[!h]
\includegraphics[width= 0.45\textwidth]{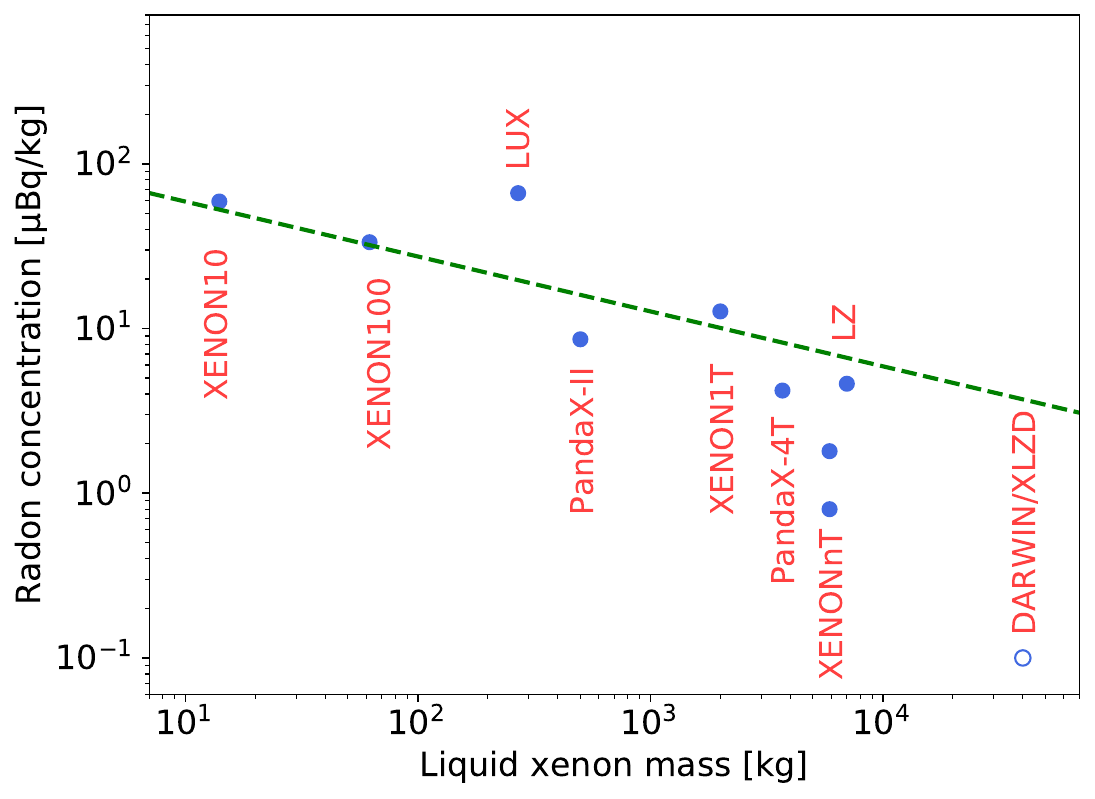}
\caption{The evolution of $^{222}$Rn concentration in two-phase Xe-TPCs (measured values, blue dots), together with the expected decrease from the surface-to-volume ratio (dashed line, $x^{-1/3}$). The goal for next-generation TPCs (open circle) is also shown.}
\label{fig:bg_radon_time}
\end{figure}

The present background goals are such that electronic and nuclear recoils rates are below the ones from irreducible astrophysical neutrino interactions.  This requirement  sets the goals for the intrinsic $^{222}$Rn and $^{85}$Kr concentrations: the background from the decay of these isotopes is to be  below the solar pp-neutrino elastic scattering rate, as shown in Figure~\ref{fig:bgs}. This condition translates into 0.1$\mu$Bq/kg for $^{222}$Rn and 0.1\,ppt for $^{\mathrm{nat}}$Kr, assuming a $^{85}$Kr/$^{\mathrm{nat}}$Kr ratio of 2$\times$10$^{-11}$.  $^{\mathrm{nat}}$Kr concentrations of $<50$\,ppq were already achieved by cryogenic distillation~\citep{XENON:2021fkt}, while for $^{222}$Rn a factor of about 10 reduction compared to the current value of 0.8\,$\mu$Bq/kg in XENONnT is  needed for future detectors, see Figure~\ref{fig:bg_radon_time}.  As an example, a $^{222}$Rn concentration of  0.1$\mu$Bq/kg corresponds to less than one radon atom per 100 mol of xenon. The main background is due to $^{214}$Bi $\beta$-decays which are not accompanied by an $\alpha$-decay and thus cannot be tagged in the TPC.

The neutron-induced NR backgrounds must be below the rate from coherent elastic neutrino-nucleus scatters (CE$\nu$Ns) from solar and atmospheric neutrinos. Muon-induced neutrons (cosmogenic)  are suppressed by going deep underground and surrounding the experiment with a large water Cherenkov shield. Neutrons from ($\alpha$,n) and fission reactions in detector components are reduced by severe material selection criteria in terms of radio-purity, and via dedicated neutron shields surrounding the cryostat (e.g., Gd-doped water or liquid scintillator). The neutron shields tag neutrons which might scatter once in the TPC and then escape from the inner detector. They also provide an {\it{in situ}} measurement of the neutron background close to the TPC.

The scattering of $^{8}$B solar neutrinos can mimic WIMPs with masses around 5-6 GeV, while neutrinos from the atmosphere and diffuse supernova neutrinos can mimic a WIMP-signal for masses above 10\,GeV. While these neutrinos will present irreducible backgrounds for the DM search, they can, on the other hand, also be promoted from backgrounds to signals, allowing to address open questions in neutrino and solar physics~\citep{Baudis:2013qla,DARWIN:2020bnc}.

\section{Overview of past, current and future detectors}
\label{sec:overview}

The first dual-phase Xe-TPCs that set competitive constraints on WIMP scatters off nuclei were those of the ZEPLIN and XENON programmes, in particular ZEPLIN-II and ZEPLIN-III at the Boulby Mine in the UK, and XENON10 at LNGS in Italy. These evolved into LUX and LUX-ZEPLIN at SURF, USA, and XENON100, XENON1T and XENONnT at LNGS. In parallel, PandaX-I and PandaX-II were constructed at the China Jinping Underground Laboratory (CJPL), followed by PandaX-4T.  Starting with total masses at the few kilogram and later 100\,kg scale, the detectors evolved and reached target masses at the tonne- and more recently multi-tonne scale. Concomitantly, the background levels in the most inner regions constantly decreased, with now unprecedented electronic recoil levels around 15 events/(t\,y\,keV) in the energy region below 100\,keV.  

The current generation of detectors employ several tons of LXe: LZ~\citep{LZ:2019sgr}, PandaX-4T~\citep{PandaX-4T:2021bab} and XENONnT~\citep{XENON:2020kmp} have total (target) LXe masses of 10\,t (7\,t), 5.6\, (3.7\,t) and 8.6\,t (5.9\,t), respectively. Their overall TPC design is rather similar, with cylindrical, PTFE enclosed target regions viewed by two arrays of 3-inch diameter Hamamatsu R11410 PMTs. The LUX-ZEPLIN and XENON programmes are reviewed by D. Akerib and E. Aprile, respectively, in this issues. All three experiments presented first results on WIMPs, as well as on other DM candidates, from their early science runs, and continue to acquire data towards their design exposures and sensitivities. No evidence for DM was found, the data being consistent with background-only hypotheses.   The minima of the spin-independent  WIMP-nucleon cross sections for the three ongoing experiments are shown in  Table~\ref{tab:constraints}. The search for WIMP dark matter is  ongoing, with a sensitivity goal around 1.5$\times$10$^{-48}$\,cm$^2$ at 40-50\,GeV/c$^2$ mass~\citep{LZ:2019sgr,XENON:2020kmp}.

\begin{table}
\centering
\begin{tabular}{l c c c} 
 \hline
 Experiment & Cross section  &  WIMP mass  \\ 
   & [cm$^2$]    & [GeV/c$^2$]         \\
   \hline
LUX-ZEPLIN & 6.5$\times$10$^{-48}$ & 30   \\ 
PandaX-4T   & 3.8$\times$10$^{-47}$  & 40  \\ 
XENONnT &    2.6$\times$10$^{-47}$ & 28  \\ 
 \hline
\end{tabular}
\caption{Minima of the upper limits on the spin-independent  WIMP-nucleon cross sections from the first science runs of the three ongoing multi-tonne xenon experiments LUX-ZEPLIN~\citep{LZ:2022ufs}, PandaX-4T~\citep{PandaX-4T:2021bab} and XENONnT~ \citep{XENON:2023sxq} .}
\label{tab:constraints}
\end{table}

The next-generation projects are DARWIN/XLZD, described in more detail in the following sections, and PandaX-xT.  DARWIN, first proposed around 2011~\citep{Baudis:2012bc}, would operate a TPC with 40\,t of LXe in the active region (50\,t in total)~\citep{Aalbers:2016jon}. In June 2021 the LZ, XENON and DARWIN collaborations signed an  MoU and joined forces to form the XLZD consortium~\citep{xlzd-website}, with the goal of constructing and operating the next-generation experiment together.  The size and scope of the detector might be enlarged, compared to DARWIN, with a 3\,m$\times$3\,m TPC containing 60\,t of LXe (75\,t in total).  PandaX-xT is the next step in the PandaX programme at CJPL, with 43\,t of LXe target in the TPC (47\,t of LXe in total). Two arrays of Hamamatsu R12699 2-inch PMTs will view the Xe volume. Compared to the 3-inch tubes employed in current TPCs, these new sensors have the advantage of lower radioactivity, faster time response and the possibility of multi-anode readout, with four independent channels per unit. The inner  cryostat vessel will be made of ultra-pure copper, and the space between the inner and outer vessel will contain an active veto. PandaX-xT aims for a 200\,t\,y exposure for WIMPs, and, similar to DARWIN/XLZD, for a broad science reach~\citep{Wang:2023wrr}.

\section{The DARWIN/XLZD Project}
\label{sec:darwin}

\begin{figure}[!h]
\centering
\includegraphics[width=0.40\textwidth]{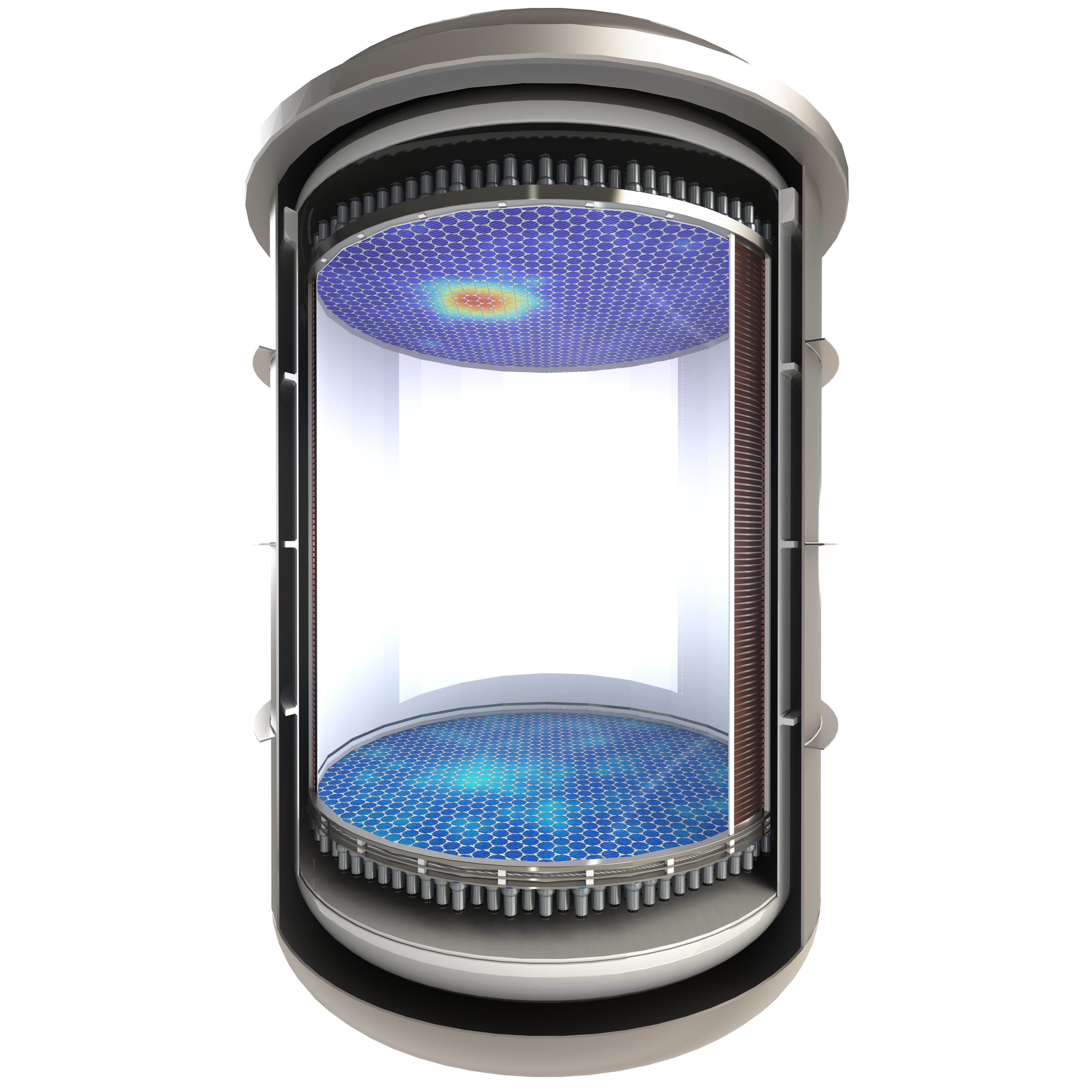}
\caption{Schematic view of the DARWIN baseline design. The cylindrical TPC, enclosed by a double-walled titanium cryostat, is 2.6\,m in diameter and height. Two arrays of  3-inch PMTs are located at the top and bottom of the TPC, which is lined with high-reflectivity PTFE and surrounded by 92 Cu field shaping rings~\citep{Aalbers:2016jon}.}
\label{fig:darwin-tpc}
\end{figure}

In its baseline design, shown in  Figure~\ref{fig:darwin-tpc}, the DARWIN experiment features a cylindrical TPC, with 2.6\,m diameter and 2.6\,m height, placed in a low-background, double-walled titanium cryostat. The cryostat and its support structure is surrounded by active neutron and muon vetos. Two photosensor arrays with a total of 1910 3-inch PMTs are located at the top and bottom of the TPC, which is lined with high-reflectivity polytetrafluoroethylene (PTFE), surrounded by 92 copper field shaping rings~\citep{Aalbers:2016jon}.  The larger mass of XLZD would imply a TPC with 3\,m diameter and 3\,m height, with a first phase employing a full-scale diameter but shallower TPC (with 40\,t active mass), and possibly a final phase with an enlarged, taller TPC, to accommodate 80\,t of active mass. 

Regardless of the exact final dimensions, a detector at the DARWIN/XLZD scale (2.6-3.0\,m) poses several technological challenges. The stringent background goals regarding radon levels require cryogenic distillation with a high liquid xenon throughput (close to 1\,t/hour) with efficient cooling power based on cryogenic heat pumps and radon-free heat exchangers.  Cryogenic distillation alone is not sufficient: it must go hand in hand with selection of materials with low radon emanation rates and, in some cases, with new coating techniques to prevent radon emanation from cryostat surfaces. Other challenges related to the liquid target are the continuous purification for electronegative impurities and water, to maintain high charge and light yields, as well as new solutions for reliable xenon storage and recuperation at large scales.  Liquid phase purification powered by a liquid xenon pump, as demonstrated in~\citep{Plante:2022khm}, was employed to achieve an electron drift lifetime above 10\,ms in about 8.6\,t of xenon in XENONnT.  Hence liquid phase purification will also be used for DARWIN, along with purification of the gaseous volume of the cryostat. The latter remains crucially important to extract radon and other impurities from detector parts with elevated impurity concentrations, such as, e.g., pipes containing cabling.

Regarding the detector design, electrodes with large diameters, high transparency, minimal sagging and low spurious electron emission, as well as high-voltage feed-throughs that can safely deliver 50\,kV or more to the cathode must be developed. The LZ collaboration successfully built custom-woven wire-mesh grids with 1.5\,m diameter, produced  with an in-house built loom to weave the wire meshes~\citep{Stifter:2020ktw}.  However, the new electrodes require a scale-up of a factor of two in linear dimensions compared to existing electrodes in XENONnT and LZ. To this end, several types of large-scale electrodes must be produced and the performance of prototypes must be assessed in test platforms above and below ground. The latter in particular would also allow to probe spurious charge and light emissions, which can contribute to the combinatorial background in the TPC, and affect the backgrounds at low energies.

The baseline TPC design will employ two arrays of low-radioactivity, 3-inch diameter PMTs (Hamamatsu R11410-21/22), as developed by Hamamatsu with the XENON and LZ collaborations over many years for operation in liquid xenon, and optimised for low radioactivity, low spurious light emission and vacuum tightness at low temperatures~\citep{Baudis:2013xva, Barrow:2016doe, Antochi:2021wik}. These PMTs have a low dark count rate ($\sim$0.02 Hz/mm$^2$) and a  high quantum efficiency ($\sim$34\%) to the xenon 175\,nm scintillation light. The arrangement of the tubes in the arrays will optimise the light collection efficiency and the $x-y$-position reconstruction using the S2-signal. While specific. activities of  $<$13.3 mBq/PMT and $<$0.6 mBq/PMT, for $^{238}$U and $^{232}$Th  were achieved, respectively, in the current generation of tubes~\citep{XENON:2015ara}, further optimisation of the employed materials will be necessary to reach the background goals.
Finally the cryostat design will be optimised to ensure stability, while reducing as much as possible the amount of material, and thus gamma and neutron emitters in proximity to the TPC.

Although the baseline detector design is well-established, and PMTs are by now a proven technology for cryogenic operation, R\&D for new type of photosensors, which could potentially replace the 3-inch tubes in future upgrades of the detector, as well as for new TPC designs, is ongoing. The studied sensors in the DARWIN collaboration include VUV-sensitive silicon photomultipliers~\citep{Sakamoto:2023ond,Peres:2023nbw,Baudis:2020nwe,Baudis:2018pdv}  and digital SiPMs, the 2-inch$\times$2-inch flat panel PMTs (Hamamatsu R12699), hybrid photosensors~\citep{DAndrea:2021fro}, as well as bubble-assisted liquid hole multipliers~\citep{Breskin:2022asf}. 

\section{Large-scale demonstrators}
\label{sec:demos}

\begin{figure}[!h]
\centering
\includegraphics[width=0.40\textwidth]{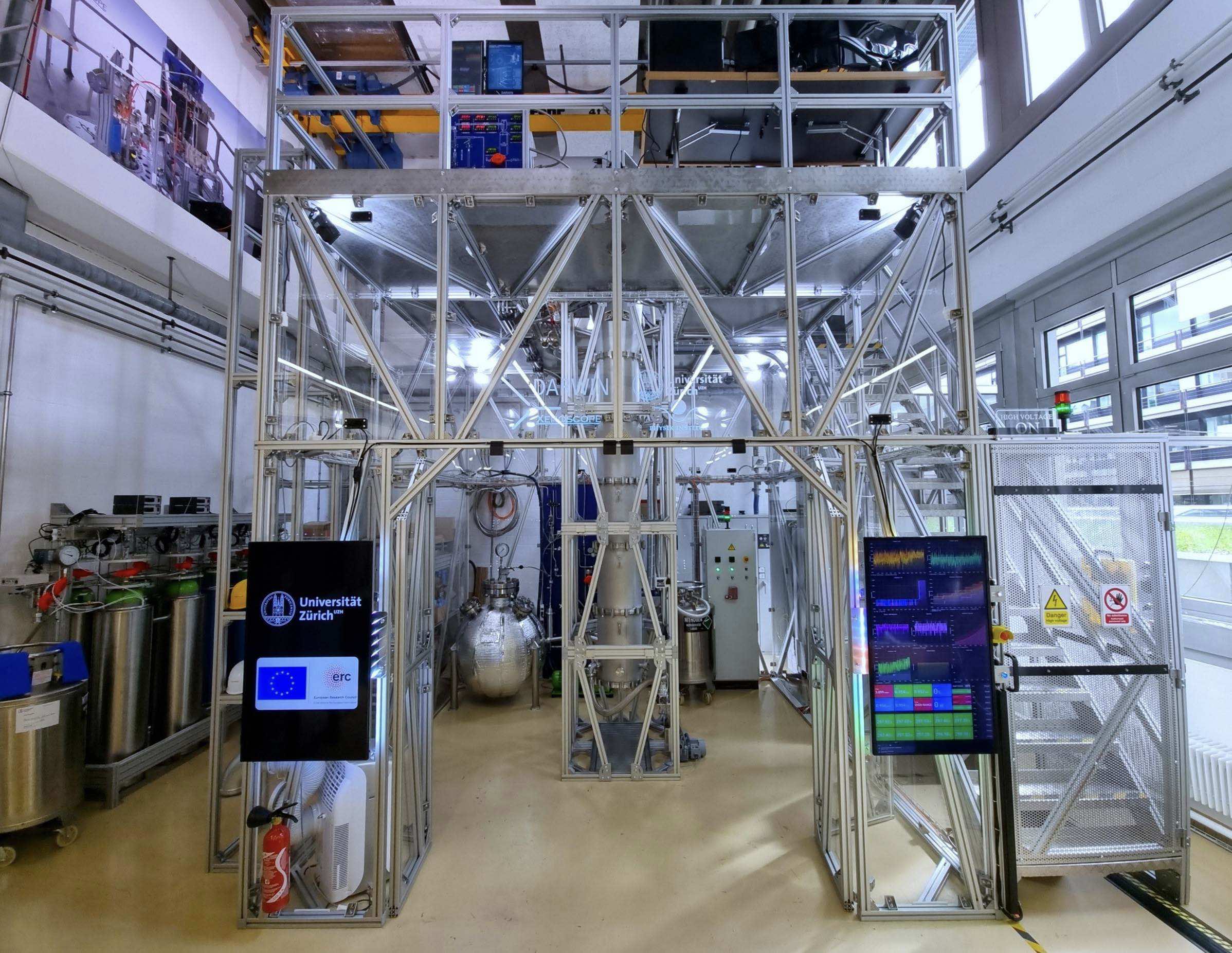}
\includegraphics[width=0.40\textwidth]{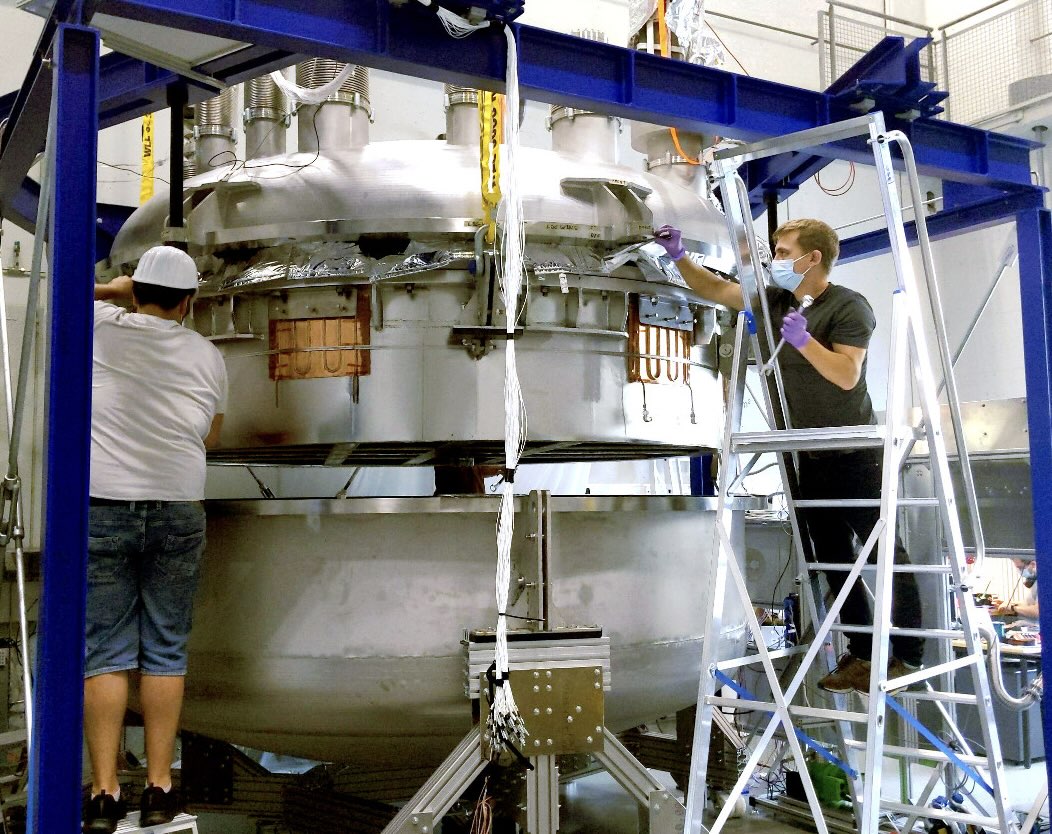}
\caption{Pictures of the large-scale DARWIN R\&D facilities Xenoscope, in Zurich (top), and Pancake, in Freiburg (bottom). Xenoscope~\citep{Baudis:2021ipf,Baudis:2023ywo} will operate a 2.6\,m tall TPC inside the 3.5\,m tall cryostat, while  Pancake will test 2.6\, diameter electrodes as well as a 2.6\, diameter TPC.}
\label{fig:darwin-demos}
\end{figure}

To address some of the challenges related to the construction and operation of such a next-generation Xe-TPC, several large-scale demonstrators have been built and are in operation. The Xenoscope facility in Zurich includes a 2.6\,m tall TPC~\citep{Baudis:2021ipf,Baudis:2023ywo}, while the Pancake facility in Freiburg allows to deploy a shallow, 2.6\, diameter TPC. The facilities, which use about 400\,kg of liquid xenon in their cryostats, are shown in Figure~\ref{fig:darwin-demos}.

\begin{figure}[!h]
\centering
\includegraphics[width=0.40\textwidth]{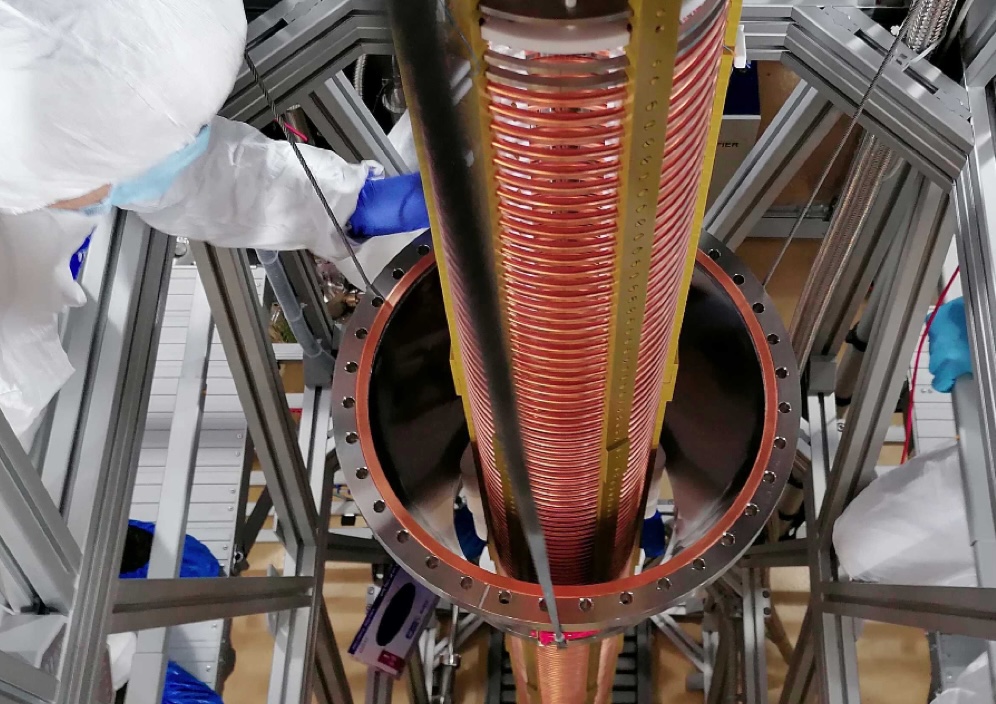}
\includegraphics[width=0.40\textwidth]{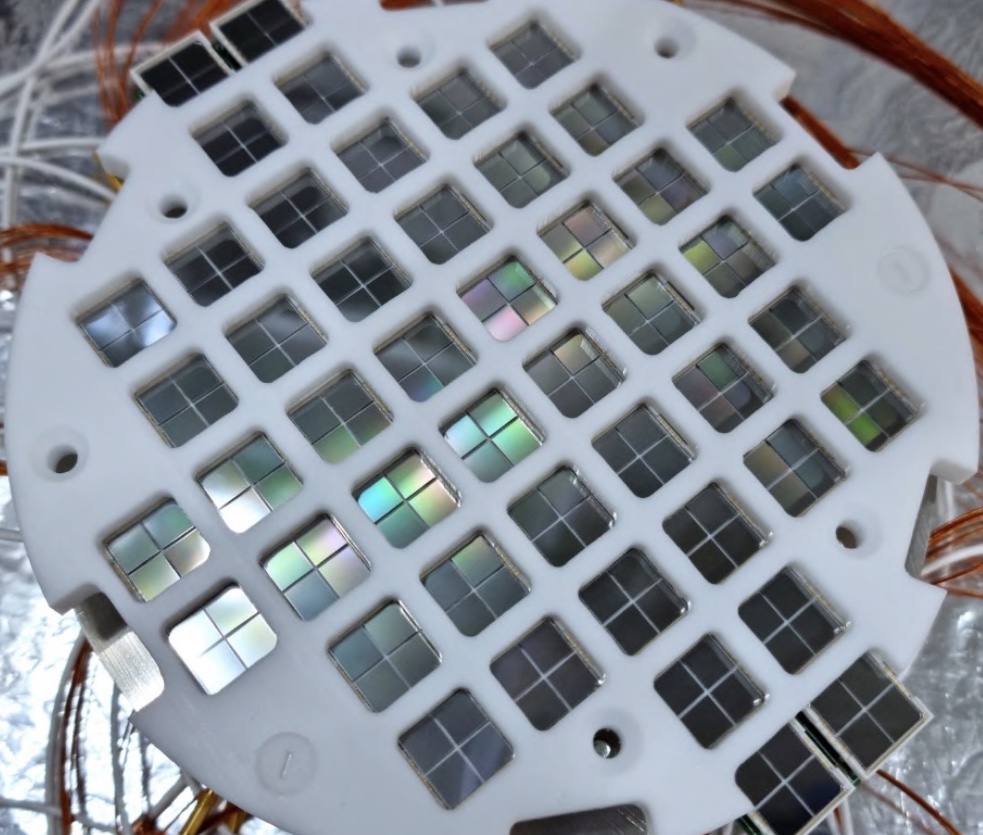}
\caption{Picture of the 2.6\,m tall TPC installed in Xenoscope (top). Visible are the Cu field shaping rings, as well as the Torlon support pillars. At the top of the TPC, a SiPM array (bottom) records the S2 signals. It is composed of twelve tiles of 24$\times$24\,mm$^2$ total area, containing 192 individual 6$\times$6\,mm$^2$ SiPM units~\citep{Peres:2023nbw}.}
\label{fig:darwin-xenoscope-tpc}
\end{figure}

The main aim of Xenoscope is to demonstrate electron drift over 2.6\,m, to measure longitudinal and transversal electron cloud diffusion for different electric drift fields, as well as light attenuation in liquid xenon over these large distances. Figure~\ref{fig:darwin-xenoscope-tpc} shows the 2.6\,m tall TPC with its top SiPM array. The main aim of Pancake is to test electrodes with large ($>$2.5\,m)  diameters. Both facilities are available as R\&D platforms to the collaboration. A new facility, LowRad, to demonstrate large-scale cryogenic distillation is under construction in M\"unster.
Apart from these large-scale demonstrators, smaller projects to study new types of inner detectors and to test new photosensors are ongoing. The new detector designs include single-phase TPCs with S2 amplification in the liquid, liquid xenon proportional counters~\citep{Qi:2023bof}, and so-called hermetic or sealed TPCs~\citep{Sato:2019qpr,Wei:2020cwl,Dierle:2022zzh}. For the latter, the inner liquid xenon volume is mechanically isolated from the rest of the detector, which contains a large fraction of radon-emanating surfaces, to prevent radon diffusion into the  sensitive xenon volume.

\section{Sensitivity to Dark Matter}
\label{sec:wimps}

With its large target mass and ultra-low background rates, the next-generation xenon detector will be a true observatory in astroparticle physics. It will not only search for dark matter, but also detect low-energy astrophysical neutrinos, search for neutrinoless double beta decay in $^{136}$Xe without the need of target enrichment, look for solar axions, as well as for other rare interactions. The science potential of a large, dual-phase xenon detector is detailed in~\citep{Aalbers:2022dzr}, a white paper signed by 600 authors. Here we briefly discuss the sensitivity to dark matter.

\begin{figure}[!h]
\centering
\includegraphics[width=0.40\textwidth]{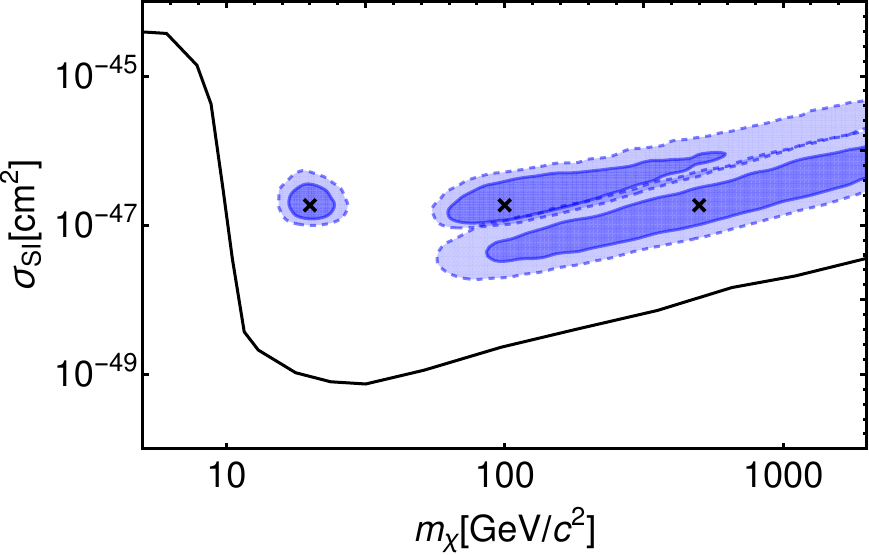}
\caption{Capability of a liquid xenon detector such as DARWIN/XLZD to reconstruct the DM mass and cross section, for three different masses, 20\,GeV, 100\,GeV and 500\,GeV and a SI WIM-nucleon cross section of 1$\times$10$^{-47}$cm$^2$. Figure from~\citep{Newstead:2013pea}.}
\label{fig:crosssection-mass}
\end{figure}

The primary science goal of DARWIN/XLZD is to discover WIMPs, or, should a first signal appear in current-generation detectors, to constrain their mass and cross section with higher sensitivity.  As an example, Figure~\ref{fig:crosssection-mass} shows the capability of a liquid xenon detector to reconstruct the DM mass and cross section, for three different masses (20\,GeV, 100\,GeV and 250\,GeV) and a SI WIMP-nucleon cross section of 1$\times$10$^{-47}$cm$^2$. The 1-$\sigma$ and 2-$\sigma$ credible regions are obtained after marginalising the posterior probability distribution over astrophysical parameters: galactic escape velocity $\mathrm{v}_{\mathrm{esc}} = (544\pm40$)\,km/s,   circular velocity $\mathrm{v}_{0} = (220\pm20$)\,km/s and  local DM density $\rho_{0} = (0.3\pm0.1$)\,GeV/cm$^3$~\citep{Newstead:2013pea}.

\begin{figure}[!h]
\centering
\includegraphics[width=0.45\textwidth]{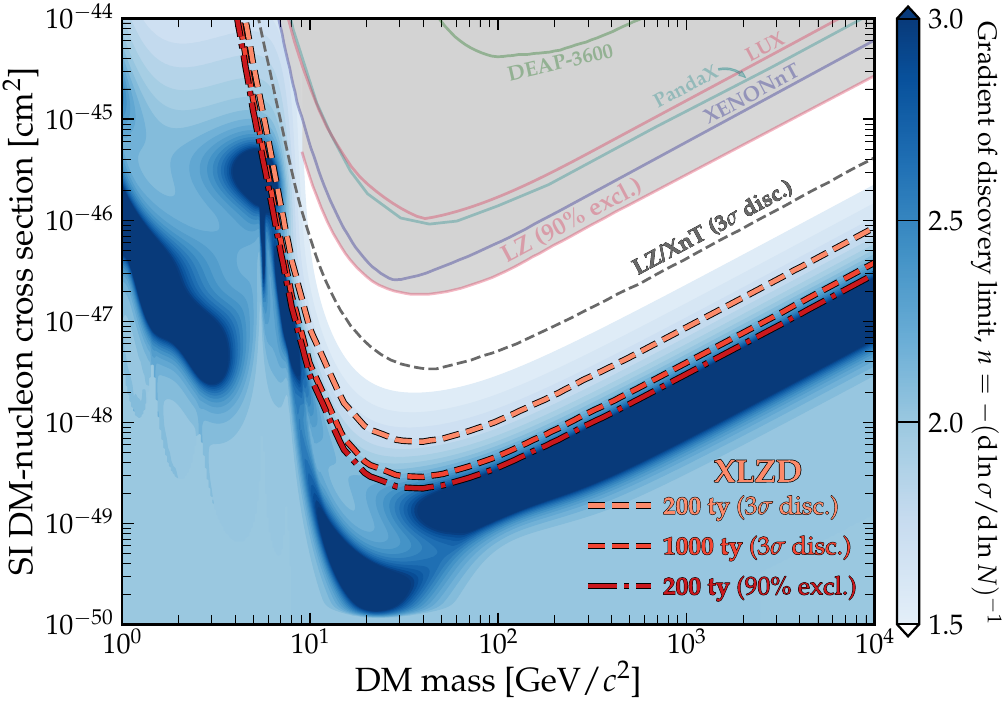}
\caption{The 3-$\sigma$ WIMP discovery sensitivity of DARWIN/XLZD at a SI cross section of 3$\times$10$^{-49}$\,cm$^2$ at 40\,GeV/c$^2$  for two different exposures, 200\,t\,y and 1000\,t\,y, as well as the 90\% exclusion sensitivity for an exposure of 200\,t\,y. The systematic limit imposed by coherent elastic neutrino nucleus scatters from solar and atmospheric neutrinos is also shown. At a given contour {\it{n}}, an increase in exposure by at least  a factor of 10$^n$  is required to probe a 10 times lower cross section~\citep{OHare:2021utq}. Figure by Ciaran O'Hare.}
\label{fig:xlzd-projections}
\end{figure}

With an exposure of 200\,t\,y, the next-generation detector will close the gap to the so-called neutrino fog, when astrophysical neutrinos will start to limit the sensitivity to WIMPs. The larger mass of XLZD, compared to the baseline DARWIN design, would allow for a 3-$\sigma$ WIMP discovery at a SI cross section of 3$\times$10$^{-49}$\,cm$^2$ at 40\,GeV/c$^2$ mass. This is illustrated in Figure~\ref{fig:xlzd-projections}, which also shows the systematic limit imposed by coherent elastic neutrino nucleus scatters from solar and atmospheric neutrinos. At a given contour {\it{n}}, an increase in exposure by at least  a factor of 10$^n$  is required to probe a 10 times lower cross section~\citep{OHare:2021utq}.
 
As already demonstrated by previous and the current generation of detectors, xenon TPCs can also search for non-WIMP dark matter candidates. These include  searches for keV-scale axion-like-particles (ALPs) and dark photons via absorption in LXe, for DM candidates in the mass range $\sim$50\,MeV-10\,GeV from a hidden sector, via DM electron scattering, and even for Planck-scale dark matter~\citep{XENON:2023iku}. DARWIN/XLZD, with a 40-60\,t LXe target, and a background dominated by neutrinos, will vastly improve upon current constraints.

\section{Summary and conclusions}
\label{sec:summary}

After decades of intense research, the fundamental nature of dark matter in the Universe remains an enigma. In the race to discover DM particles, dual-phase Xe-TPCs remain at the forefront.  Developed in the early twenty-first century primarily to search for dark matter in the form of WIMPs, they soon surpassed other technologies in terms of their sensitivity to WIMP-nucleon interactions over a large range of WIMP masses~\citep{Baudis:2012ig}. Almost twenty years later, detectors using several tonnes of liquid xenon can observe signals down to a few quanta (photons and electrons) with unprecedented low background rates, approaching the irreducible background from astrophysical neutrinos.  While the current generation of detectors continue to acquire data in different deep underground laboratories, a next-generation experiments at the multi-ten-ton scale, DARWIN/XLZD,  is in planning.  Although by now an established technology, the scaling up of the Xe-TPCs from diameters and heights of 1.5\,m to about twice these dimensions poses several technological challenges. To address these, large-scale R\&D projects are ongoing, with two full-scale demonstrators in operation.  Several smaller scale detectors explore alternative designs and photosensors, which, albeit not part of the baseline design, could be potentially employed in future upgrades of the detector. With foreseen first data in the early 2030s, DARWIN/XLZD might discover dark matter particles, and thus solve a problem which is almost a century old. At the same time, it will break new grounds in other areas of astroparticle physics, in particular in neutrino physics and very rare nuclear transitions.

\section*{Acknowledgements}
I thank the organisers for a very stimulating symposium in beautiful surroundings. I acknowledge support from the University of Zurich, from the European Research Council (ERC) under the European Union's Horizon 2020 research and innovation programme, grant agreement No. 742789 ({\sl Xenoscope}) and from the SNF via grants 200020-219290 and 20FL20-216573.

\bibliographystyle{elsarticle-harv}
\bibliography{baudis_dm}

\begin{thebibliography}{38}
\expandafter\ifx\csname natexlab\endcsname\relax\def\natexlab#1{#1}\fi
\providecommand{\url}[1]{\texttt{#1}}
\providecommand{\href}[2]{#2}
\providecommand{\path}[1]{#1}
\providecommand{\DOIprefix}{doi:}
\providecommand{\ArXivprefix}{arXiv:}
\providecommand{\URLprefix}{URL: }
\providecommand{\Pubmedprefix}{pmid:}
\providecommand{\doi}[1]{\href{http://dx.doi.org/#1}{\path{#1}}}
\providecommand{\Pubmed}[1]{\href{pmid:#1}{\path{#1}}}
\providecommand{\bibinfo}[2]{#2}
\ifx\xfnm\relax \def\xfnm[#1]{\unskip,\space#1}\fi
\bibitem[{Aalbers et~al.(2016)}]{Aalbers:2016jon}
\bibinfo{author}{Aalbers, J.}, et~al. (\bibinfo{collaboration}{DARWIN}),
  \bibinfo{year}{2016}.
\newblock \bibinfo{title}{{DARWIN: towards the ultimate dark matter detector}}.
\newblock \bibinfo{journal}{JCAP} \bibinfo{volume}{1611}, \bibinfo{pages}{017}.
\newblock \DOIprefix\doi{10.1088/1475-7516/2016/11/017},
  \href{http://arxiv.org/abs/1606.07001}{{\tt arXiv:1606.07001}}.
\bibitem[{Aalbers et~al.(2020)}]{DARWIN:2020bnc}
\bibinfo{author}{Aalbers, J.}, et~al. (\bibinfo{collaboration}{DARWIN}),
  \bibinfo{year}{2020}.
\newblock \bibinfo{title}{{Solar neutrino detection sensitivity in DARWIN via
  electron scattering}}.
\newblock \bibinfo{journal}{Eur. Phys. J. C} \bibinfo{volume}{80},
  \bibinfo{pages}{1133}.
\newblock \DOIprefix\doi{10.1140/epjc/s10052-020-08602-7},
  \href{http://arxiv.org/abs/2006.03114}{{\tt arXiv:2006.03114}}.
\bibitem[{Aalbers et~al.(2023a)}]{Aalbers:2022dzr}
\bibinfo{author}{Aalbers, J.}, et~al., \bibinfo{year}{2023}a.
\newblock \bibinfo{title}{{A next-generation liquid xenon observatory for dark
  matter and neutrino physics}}.
\newblock \bibinfo{journal}{J. Phys. G} \bibinfo{volume}{50},
  \bibinfo{pages}{013001}.
\newblock \DOIprefix\doi{10.1088/1361-6471/ac841a},
  \href{http://arxiv.org/abs/2203.02309}{{\tt arXiv:2203.02309}}.
\bibitem[{Aalbers et~al.(2023b)}]{LZ:2022ufs}
\bibinfo{author}{Aalbers, J.}, et~al. (\bibinfo{collaboration}{LZ}),
  \bibinfo{year}{2023}b.
\newblock \bibinfo{title}{{First Dark Matter Search Results from the LUX-ZEPLIN
  (LZ) Experiment}}.
\newblock \bibinfo{journal}{Phys. Rev. Lett.} \bibinfo{volume}{131},
  \bibinfo{pages}{041002}.
\newblock \DOIprefix\doi{https://doi.org/10.1103/PhysRevLett.131.041002},
  \href{http://arxiv.org/abs/2207.03764}{{\tt arXiv:2207.03764}}.
\bibitem[{Aghanim et~al.(2020)}]{Planck:2018vyg}
\bibinfo{author}{Aghanim, N.}, et~al. (\bibinfo{collaboration}{Planck}),
  \bibinfo{year}{2020}.
\newblock \bibinfo{title}{{Planck 2018 results. VI. Cosmological parameters}}.
\newblock \bibinfo{journal}{Astron. Astrophys.} \bibinfo{volume}{641},
  \bibinfo{pages}{A6}.
\newblock \DOIprefix\doi{10.1051/0004-6361/201833910},
  \href{http://arxiv.org/abs/1807.06209}{{\tt arXiv:1807.06209}}.
  \bibinfo{note}{[Erratum: Astron.Astrophys. 652, C4 (2021)]}.
\bibitem[{Akerib et~al.(2020)}]{LZ:2019sgr}
\bibinfo{author}{Akerib, D.S.}, et~al. (\bibinfo{collaboration}{LZ}),
  \bibinfo{year}{2020}.
\newblock \bibinfo{title}{{The LUX-ZEPLIN (LZ) Experiment}}.
\newblock \bibinfo{journal}{Nucl. Instrum. Meth. A} \bibinfo{volume}{953},
  \bibinfo{pages}{163047}.
\newblock \DOIprefix\doi{10.1016/j.nima.2019.163047},
  \href{http://arxiv.org/abs/1910.09124}{{\tt arXiv:1910.09124}}.
\bibitem[{Antochi et~al.(2021)}]{Antochi:2021wik}
\bibinfo{author}{Antochi, V.C.}, et~al., \bibinfo{year}{2021}.
\newblock \bibinfo{title}{{Improved quality tests of R11410-21 photomultiplier
  tubes for the XENONnT experiment}}.
\newblock \bibinfo{journal}{JINST} \bibinfo{volume}{16},
  \bibinfo{pages}{P08033}.
\newblock \DOIprefix\doi{10.1088/1748-0221/16/08/P08033},
  \href{http://arxiv.org/abs/2104.15051}{{\tt arXiv:2104.15051}}.
\bibitem[{Aprile et~al.(2015)}]{XENON:2015ara}
\bibinfo{author}{Aprile, E.}, et~al. (\bibinfo{collaboration}{XENON}),
  \bibinfo{year}{2015}.
\newblock \bibinfo{title}{{Lowering the radioactivity of the photomultiplier
  tubes for the XENON1T dark matter experiment}}.
\newblock \bibinfo{journal}{Eur. Phys. J. C} \bibinfo{volume}{75},
  \bibinfo{pages}{546}.
\newblock \DOIprefix\doi{10.1140/epjc/s10052-015-3657-5},
  \href{http://arxiv.org/abs/1503.07698}{{\tt arXiv:1503.07698}}.
\bibitem[{Aprile et~al.(2020)}]{XENON:2020kmp}
\bibinfo{author}{Aprile, E.}, et~al. (\bibinfo{collaboration}{XENON}),
  \bibinfo{year}{2020}.
\newblock \bibinfo{title}{{Projected WIMP sensitivity of the XENONnT dark
  matter experiment}}.
\newblock \bibinfo{journal}{JCAP} \bibinfo{volume}{11}, \bibinfo{pages}{031}.
\newblock \DOIprefix\doi{10.1088/1475-7516/2020/11/031},
  \href{http://arxiv.org/abs/2007.08796}{{\tt arXiv:2007.08796}}.
\bibitem[{Aprile et~al.(2022)}]{XENON:2021fkt}
\bibinfo{author}{Aprile, E.}, et~al. (\bibinfo{collaboration}{XENON}),
  \bibinfo{year}{2022}.
\newblock \bibinfo{title}{{Application and modeling of an online distillation
  method to reduce krypton and argon in XENON1T}}.
\newblock \bibinfo{journal}{PTEP} \bibinfo{volume}{2022},
  \bibinfo{pages}{053H01}.
\newblock \DOIprefix\doi{10.1093/ptep/ptac074},
  \href{http://arxiv.org/abs/2112.12231}{{\tt arXiv:2112.12231}}.
\bibitem[{Aprile et~al.(2023a)}]{XENON:2023sxq}
\bibinfo{author}{Aprile, E.}, et~al. (\bibinfo{collaboration}{XENON}),
  \bibinfo{year}{2023}a.
\newblock \bibinfo{title}{{First Dark Matter Search with Nuclear Recoils from
  the XENONnT Experiment}}.
\newblock \bibinfo{journal}{Phys. Rev. Lett.} \bibinfo{volume}{131},
  \bibinfo{pages}{041003}.
\newblock \DOIprefix\doi{10.1103/PhysRevLett.131.041003},
  \href{http://arxiv.org/abs/2303.14729}{{\tt arXiv:2303.14729}}.
\bibitem[{Aprile et~al.(2023b)}]{XENON:2023iku}
\bibinfo{author}{Aprile, E.}, et~al. (\bibinfo{collaboration}{XENON}),
  \bibinfo{year}{2023}b.
\newblock \bibinfo{title}{{Searching for Heavy Dark Matter near the Planck Mass
  with XENON1T}}.
\newblock \bibinfo{journal}{Phys. Rev. Lett.} \bibinfo{volume}{130},
  \bibinfo{pages}{261002}.
\newblock \DOIprefix\doi{10.1103/PhysRevLett.130.261002},
  \href{http://arxiv.org/abs/2304.10931}{{\tt arXiv:2304.10931}}.
\bibitem[{Barrow et~al.(2017)}]{Barrow:2016doe}
\bibinfo{author}{Barrow, P.}, et~al., \bibinfo{year}{2017}.
\newblock \bibinfo{title}{{Qualification Tests of the R11410-21 Photomultiplier
  Tubes for the XENON1T Detector}}.
\newblock \bibinfo{journal}{JINST} \bibinfo{volume}{12},
  \bibinfo{pages}{P01024}.
\newblock \DOIprefix\doi{10.1088/1748-0221/12/01/P01024},
  \href{http://arxiv.org/abs/1609.01654}{{\tt arXiv:1609.01654}}.
\bibitem[{Baudis(2012a)}]{Baudis:2012bc}
\bibinfo{author}{Baudis, L.} (\bibinfo{collaboration}{DARWIN Consortium}),
  \bibinfo{year}{2012}a.
\newblock \bibinfo{title}{{DARWIN: dark matter WIMP search with noble
  liquids}}.
\newblock \bibinfo{journal}{J. Phys. Conf. Ser.} \bibinfo{volume}{375},
  \bibinfo{pages}{012028}.
\newblock \DOIprefix\doi{10.1088/1742-6596/375/1/012028},
  \href{http://arxiv.org/abs/1201.2402}{{\tt arXiv:1201.2402}}.
\bibitem[{Baudis(2012b)}]{Baudis:2012ig}
\bibinfo{author}{Baudis, L.}, \bibinfo{year}{2012}b.
\newblock \bibinfo{title}{{Direct dark matter detection: the next decade}}.
\newblock \bibinfo{journal}{Phys. Dark Univ.} \bibinfo{volume}{1},
  \bibinfo{pages}{94--108}.
\newblock \DOIprefix\doi{10.1016/j.dark.2012.10.006},
  \href{http://arxiv.org/abs/1211.7222}{{\tt arXiv:1211.7222}}.
\bibitem[{Baudis(2023)}]{Baudis:2023pzu}
\bibinfo{author}{Baudis, L.}, \bibinfo{year}{2023}.
\newblock \bibinfo{title}{{Dual-phase xenon time projection chambers for
  rare-event searches}}.
\newblock \bibinfo{journal}{Phil.Trans. R. Soc. A} \bibinfo{volume}{382},
  \bibinfo{pages}{20230083}.
\newblock \DOIprefix\doi{10.1098/rsta.2023.0083},
  \href{http://arxiv.org/abs/2311.05320}{{\tt arXiv:2311.05320}}.
\bibitem[{Baudis et~al.(2013)Baudis, Behrens, Ferella, Kish,
  Marrodan~Undagoitia, Mayani and Schumann}]{Baudis:2013xva}
\bibinfo{author}{Baudis, L.}, \bibinfo{author}{Behrens, A.},
  \bibinfo{author}{Ferella, A.}, \bibinfo{author}{Kish, A.},
  \bibinfo{author}{Marrodan~Undagoitia, T.}, \bibinfo{author}{Mayani, D.},
  \bibinfo{author}{Schumann, M.}, \bibinfo{year}{2013}.
\newblock \bibinfo{title}{{Performance of the Hamamatsu R11410 Photomultiplier
  Tube in cryogenic Xenon Environments}}.
\newblock \bibinfo{journal}{JINST} \bibinfo{volume}{8},
  \bibinfo{pages}{P04026}.
\newblock \DOIprefix\doi{10.1088/1748-0221/8/04/P04026},
  \href{http://arxiv.org/abs/1303.0226}{{\tt arXiv:1303.0226}}.
\bibitem[{Baudis et~al.(2020)Baudis, Biondi, Galloway, Girard, Hochrein,
  Reichard, Sanchez-Lucas, Thieme and Wulf}]{Baudis:2020nwe}
\bibinfo{author}{Baudis, L.}, \bibinfo{author}{Biondi, Y.},
  \bibinfo{author}{Galloway, M.}, \bibinfo{author}{Girard, F.},
  \bibinfo{author}{Hochrein, S.}, \bibinfo{author}{Reichard, S.},
  \bibinfo{author}{Sanchez-Lucas, P.}, \bibinfo{author}{Thieme, K.},
  \bibinfo{author}{Wulf, J.}, \bibinfo{year}{2020}.
\newblock \bibinfo{title}{{The first dual-phase xenon TPC equipped with silicon
  photomultipliers and characterisation with $^{37}\hbox {Ar}$}}.
\newblock \bibinfo{journal}{Eur. Phys. J. C} \bibinfo{volume}{80},
  \bibinfo{pages}{477}.
\newblock \DOIprefix\doi{10.1140/epjc/s10052-020-8031-6},
  \href{http://arxiv.org/abs/2003.01731}{{\tt arXiv:2003.01731}}.
\bibitem[{Baudis et~al.(2021)Baudis, Biondi, Galloway, Girard, Manfredini,
  McFadden, Peres, Sanchez-Lucas and Thieme}]{Baudis:2021ipf}
\bibinfo{author}{Baudis, L.}, \bibinfo{author}{Biondi, Y.},
  \bibinfo{author}{Galloway, M.}, \bibinfo{author}{Girard, F.},
  \bibinfo{author}{Manfredini, A.}, \bibinfo{author}{McFadden, N.},
  \bibinfo{author}{Peres, R.}, \bibinfo{author}{Sanchez-Lucas, P.},
  \bibinfo{author}{Thieme, K.}, \bibinfo{year}{2021}.
\newblock \bibinfo{title}{{Design and construction of Xenoscope \textemdash{} a
  full-scale vertical demonstrator for the DARWIN observatory}}.
\newblock \bibinfo{journal}{JINST} \bibinfo{volume}{16},
  \bibinfo{pages}{P08052}.
\newblock \DOIprefix\doi{10.1088/1748-0221/16/08/P08052},
  \href{http://arxiv.org/abs/2105.13829}{{\tt arXiv:2105.13829}}.
\bibitem[{Baudis et~al.(2018)Baudis, Galloway, Kish, Marentini and
  Wulf}]{Baudis:2018pdv}
\bibinfo{author}{Baudis, L.}, \bibinfo{author}{Galloway, M.},
  \bibinfo{author}{Kish, A.}, \bibinfo{author}{Marentini, C.},
  \bibinfo{author}{Wulf, J.}, \bibinfo{year}{2018}.
\newblock \bibinfo{title}{{Characterisation of Silicon Photomultipliers for
  Liquid Xenon Detectors}}.
\newblock \bibinfo{journal}{JINST} \bibinfo{volume}{13},
  \bibinfo{pages}{P10022}.
\newblock \DOIprefix\doi{10.1088/1748-0221/13/10/P10022},
  \href{http://arxiv.org/abs/1808.06827}{{\tt arXiv:1808.06827}}.
\bibitem[{Baudis et~al.(2014)}]{Baudis:2013qla}
\bibinfo{author}{Baudis, L.}, et~al., \bibinfo{year}{2014}.
\newblock \bibinfo{title}{{Neutrino physics with multi-ton scale liquid xenon
  detectors}}.
\newblock \bibinfo{journal}{JCAP} \bibinfo{volume}{01}, \bibinfo{pages}{044}.
\newblock \DOIprefix\doi{10.1088/1475-7516/2014/01/044},
  \href{http://arxiv.org/abs/1309.7024}{{\tt arXiv:1309.7024}}.
\bibitem[{Baudis et~al.(2023)}]{Baudis:2023ywo}
\bibinfo{author}{Baudis, L.}, et~al., \bibinfo{year}{2023}.
\newblock \bibinfo{title}{{Electron transport measurements in liquid xenon with
  Xenoscope, a large-scale DARWIN demonstrator}}.
\newblock \bibinfo{journal}{Eur. Phys. J. C} \bibinfo{volume}{83},
  \bibinfo{pages}{717}.
\newblock \DOIprefix\doi{10.1140/epjc/s10052-023-11823-1},
  \href{http://arxiv.org/abs/2303.13963}{{\tt arXiv:2303.13963}}.
\bibitem[{Breskin(2022)}]{Breskin:2022asf}
\bibinfo{author}{Breskin, A.}, \bibinfo{year}{2022}.
\newblock \bibinfo{title}{{Novel electron and photon recording concepts in
  noble-liquid detectors}}.
\newblock \bibinfo{journal}{JINST} \bibinfo{volume}{17},
  \bibinfo{pages}{P08002}.
\newblock \DOIprefix\doi{10.1088/1748-0221/17/08/P08002},
  \href{http://arxiv.org/abs/2203.01774}{{\tt arXiv:2203.01774}}.
\bibitem[{Consortium(2022)}]{xlzd-website}
\bibinfo{author}{Consortium, X.}, \bibinfo{year}{2022}.
\newblock \bibinfo{title}{{XLZD Dark Matter Detection Consortium}}.
\newblock \bibinfo{howpublished}{{\url {https://xlzd.org}}}.
\newblock \bibinfo{note}{Accessed: 2023-06-23}.
\bibitem[{D'Andrea et~al.(2022)D'Andrea, Biondi, Ferrari, Ferella, Mahlstedt
  and Pieramico}]{DAndrea:2021fro}
\bibinfo{author}{D'Andrea, V.}, \bibinfo{author}{Biondi, R.},
  \bibinfo{author}{Ferrari, C.}, \bibinfo{author}{Ferella, A.D.},
  \bibinfo{author}{Mahlstedt, J.}, \bibinfo{author}{Pieramico, G.},
  \bibinfo{year}{2022}.
\newblock \bibinfo{title}{{The ABALONE photosensor}}.
\newblock \bibinfo{journal}{JINST} \bibinfo{volume}{17},
  \bibinfo{pages}{C01038}.
\newblock \DOIprefix\doi{10.1088/1748-0221/17/01/C01038},
  \href{http://arxiv.org/abs/2111.02924}{{\tt arXiv:2111.02924}}.
\bibitem[{Dierle et~al.(2023)Dierle, Brown, Fischer, Glade-Beucke, Grigat,
  Kuger, Lindemann, Silva and Schumann}]{Dierle:2022zzh}
\bibinfo{author}{Dierle, J.}, \bibinfo{author}{Brown, A.},
  \bibinfo{author}{Fischer, H.}, \bibinfo{author}{Glade-Beucke, R.},
  \bibinfo{author}{Grigat, J.}, \bibinfo{author}{Kuger, F.},
  \bibinfo{author}{Lindemann, S.}, \bibinfo{author}{Silva, M.R.},
  \bibinfo{author}{Schumann, M.}, \bibinfo{year}{2023}.
\newblock \bibinfo{title}{{Reduction of $^{222}\hbox {Rn}$-induced backgrounds
  in a hermetic dual-phase xenon time projection chamber}}.
\newblock \bibinfo{journal}{Eur. Phys. J. C} \bibinfo{volume}{83},
  \bibinfo{pages}{9}.
\newblock \DOIprefix\doi{10.1140/epjc/s10052-022-11151-w},
  \href{http://arxiv.org/abs/2209.00362}{{\tt arXiv:2209.00362}}.
\bibitem[{Meng et~al.(2021)}]{PandaX-4T:2021bab}
\bibinfo{author}{Meng, Y.}, et~al. (\bibinfo{collaboration}{PandaX-4T}),
  \bibinfo{year}{2021}.
\newblock \bibinfo{title}{{Dark Matter Search Results from the PandaX-4T
  Commissioning Run}}.
\newblock \bibinfo{journal}{Phys. Rev. Lett.} \bibinfo{volume}{127},
  \bibinfo{pages}{261802}.
\newblock \DOIprefix\doi{10.1103/PhysRevLett.127.261802},
  \href{http://arxiv.org/abs/2107.13438}{{\tt arXiv:2107.13438}}.
\bibitem[{Newstead et~al.(2013)Newstead, Jacques, Krauss, Dent and
  Ferrer}]{Newstead:2013pea}
\bibinfo{author}{Newstead, J.L.}, \bibinfo{author}{Jacques, T.D.},
  \bibinfo{author}{Krauss, L.M.}, \bibinfo{author}{Dent, J.B.},
  \bibinfo{author}{Ferrer, F.}, \bibinfo{year}{2013}.
\newblock \bibinfo{title}{{Scientific reach of multiton-scale dark matter
  direct detection experiments}}.
\newblock \bibinfo{journal}{Phys. Rev. D} \bibinfo{volume}{88},
  \bibinfo{pages}{076011}.
\newblock \DOIprefix\doi{10.1103/PhysRevD.88.076011},
  \href{http://arxiv.org/abs/1306.3244}{{\tt arXiv:1306.3244}}.
\bibitem[{O'Hare(2021)}]{OHare:2021utq}
\bibinfo{author}{O'Hare, C.A.J.}, \bibinfo{year}{2021}.
\newblock \bibinfo{title}{{New Definition of the Neutrino Floor for Direct Dark
  Matter Searches}}.
\newblock \bibinfo{journal}{Phys. Rev. Lett.} \bibinfo{volume}{127},
  \bibinfo{pages}{251802}.
\newblock \DOIprefix\doi{10.1103/PhysRevLett.127.251802},
  \href{http://arxiv.org/abs/2109.03116}{{\tt arXiv:2109.03116}}.
\bibitem[{Peres(2023)}]{Peres:2023nbw}
\bibinfo{author}{Peres, R.} (\bibinfo{collaboration}{Xenoscope Team}),
  \bibinfo{year}{2023}.
\newblock \bibinfo{title}{{SiPM array of Xenoscope, a full-scale DARWIN
  vertical demonstrator}}.
\newblock \bibinfo{journal}{JINST} \bibinfo{volume}{18},
  \bibinfo{pages}{C03027}.
\newblock \DOIprefix\doi{10.1088/1748-0221/18/03/C03027},
  \href{http://arxiv.org/abs/2303.15300}{{\tt arXiv:2303.15300}}.
\bibitem[{Plante et~al.(2022)Plante, Aprile, Howlett and
  Zhang}]{Plante:2022khm}
\bibinfo{author}{Plante, G.}, \bibinfo{author}{Aprile, E.},
  \bibinfo{author}{Howlett, J.}, \bibinfo{author}{Zhang, Y.},
  \bibinfo{year}{2022}.
\newblock \bibinfo{title}{{Liquid-phase purification for multi-tonne xenon
  detectors}}.
\newblock \bibinfo{journal}{Eur. Phys. J. C} \bibinfo{volume}{82},
  \bibinfo{pages}{860}.
\newblock \DOIprefix\doi{10.1140/epjc/s10052-022-10832-w},
  \href{http://arxiv.org/abs/2205.07336}{{\tt arXiv:2205.07336}}.
\bibitem[{Qi et~al.(2023)Qi, Hood, Kopec, Ma, Xu, Zhong and Ni}]{Qi:2023bof}
\bibinfo{author}{Qi, J.}, \bibinfo{author}{Hood, N.}, \bibinfo{author}{Kopec,
  A.}, \bibinfo{author}{Ma, Y.}, \bibinfo{author}{Xu, H.},
  \bibinfo{author}{Zhong, M.}, \bibinfo{author}{Ni, K.}, \bibinfo{year}{2023}.
\newblock \bibinfo{title}{{Low energy electronic recoils and single electron
  detection with a liquid Xenon proportional scintillation counter}}.
\newblock \bibinfo{journal}{JINST} \bibinfo{volume}{18},
  \bibinfo{pages}{P07027}.
\newblock \DOIprefix\doi{10.1088/1748-0221/18/07/P07027},
  \href{http://arxiv.org/abs/2301.12296}{{\tt arXiv:2301.12296}}.
\bibitem[{Sakamoto et~al.(2023)Sakamoto, Hasegawa, Itow, Kazama, Kobayashi and
  Yamashita}]{Sakamoto:2023ond}
\bibinfo{author}{Sakamoto, S.}, \bibinfo{author}{Hasegawa, T.},
  \bibinfo{author}{Itow, Y.}, \bibinfo{author}{Kazama, S.},
  \bibinfo{author}{Kobayashi, M.}, \bibinfo{author}{Yamashita, M.},
  \bibinfo{year}{2023}.
\newblock \bibinfo{title}{{Development of a low-noise SiPM for the DARWIN
  experiment}}.
\newblock \bibinfo{journal}{PoS} \bibinfo{volume}{ICRC2023},
  \bibinfo{pages}{1435}.
\newblock \DOIprefix\doi{10.22323/1.444.1435}.
\bibitem[{Sato et~al.(2020)Sato, Yamashita, Ichimura, Itow, Kazama, Moriyama,
  Ozaki, Suzuki and Yamazaki}]{Sato:2019qpr}
\bibinfo{author}{Sato, K.}, \bibinfo{author}{Yamashita, M.},
  \bibinfo{author}{Ichimura, K.}, \bibinfo{author}{Itow, Y.},
  \bibinfo{author}{Kazama, S.}, \bibinfo{author}{Moriyama, S.},
  \bibinfo{author}{Ozaki, K.}, \bibinfo{author}{Suzuki, T.},
  \bibinfo{author}{Yamazaki, R.}, \bibinfo{year}{2020}.
\newblock \bibinfo{title}{{Development of a dual-phase xenon TPC with a quartz
  chamber for direct dark matter searches}}.
\newblock \bibinfo{journal}{PTEP} \bibinfo{volume}{2020},
  \bibinfo{pages}{113H02}.
\newblock \DOIprefix\doi{10.1093/ptep/ptaa141},
  \href{http://arxiv.org/abs/1910.13831}{{\tt arXiv:1910.13831}}.
\bibitem[{Stifter(2020)}]{Stifter:2020ktw}
\bibinfo{author}{Stifter, K.} (\bibinfo{collaboration}{LZ}),
  \bibinfo{year}{2020}.
\newblock \bibinfo{title}{{Development and performance of high voltage
  electrodes for the LZ experiment}}.
\newblock \bibinfo{journal}{J. Phys. Conf. Ser.} \bibinfo{volume}{1468},
  \bibinfo{pages}{012016}.
\newblock \DOIprefix\doi{10.1088/1742-6596/1468/1/012016}.
\bibitem[{Wang et~al.(2023)Wang, Lei, Ju, Liu, Zhou, Chen, Wang, Cui, Meng and
  Zhao}]{Wang:2023wrr}
\bibinfo{author}{Wang, X.}, \bibinfo{author}{Lei, Z.}, \bibinfo{author}{Ju,
  Y.}, \bibinfo{author}{Liu, J.}, \bibinfo{author}{Zhou, N.},
  \bibinfo{author}{Chen, Y.}, \bibinfo{author}{Wang, Z.}, \bibinfo{author}{Cui,
  X.}, \bibinfo{author}{Meng, Y.}, \bibinfo{author}{Zhao, L.},
  \bibinfo{year}{2023}.
\newblock \bibinfo{title}{{Design, construction and commissioning of the
  PandaX-30T liquid xenon management system}}.
\newblock \bibinfo{journal}{JINST} \bibinfo{volume}{18},
  \bibinfo{pages}{P05028}.
\newblock \DOIprefix\doi{10.1088/1748-0221/18/05/P05028},
  \href{http://arxiv.org/abs/2301.06044}{{\tt arXiv:2301.06044}}.
\bibitem[{Wei et~al.(2021)Wei, Long, Lombardi, Jiang, Ye and Ni}]{Wei:2020cwl}
\bibinfo{author}{Wei, Y.}, \bibinfo{author}{Long, J.},
  \bibinfo{author}{Lombardi, F.}, \bibinfo{author}{Jiang, Z.},
  \bibinfo{author}{Ye, J.}, \bibinfo{author}{Ni, K.}, \bibinfo{year}{2021}.
\newblock \bibinfo{title}{{Development and Performance of a Sealed Liquid Xenon
  Time Projection Chamber}}.
\newblock \bibinfo{journal}{JINST} \bibinfo{volume}{16},
  \bibinfo{pages}{P01018}.
\newblock \DOIprefix\doi{10.1088/1748-0221/16/01/P01018},
  \href{http://arxiv.org/abs/2007.16194}{{\tt arXiv:2007.16194}}.
\bibitem[{Workman et~al.(2022)}]{ParticleDataGroup:2022pth}
\bibinfo{author}{Workman, R.L.}, et~al. (\bibinfo{collaboration}{Particle Data
  Group}), \bibinfo{year}{2022}.
\newblock \bibinfo{title}{{Review of Particle Physics}}.
\newblock \bibinfo{journal}{PTEP} \bibinfo{volume}{2022},
  \bibinfo{pages}{083C01}.
\newblock \DOIprefix\doi{10.1093/ptep/ptac097}.

\end{thebibliography}

\end{document}